\documentclass{aa}  
\usepackage{graphicx}
\usepackage{multirow}
\usepackage[colorlinks=true, allcolors=blue]{hyperref}
\usepackage{booktabs}
\usepackage{float}
\usepackage{txfonts}
\usepackage{amsmath}
\usepackage{balance}

\defcitealias{Y26}{Y26}

\def\sgra{\object{Sgr~A*}\xspace}

\newcommand{\kc}{\Omega_{\mathrm{coef}}\xspace}
\newcommand{\rs}{r_{\mathrm{hs}}\xspace}
\newcommand{\dpa}{\Delta PA\xspace}

\newcommand\ipole{{\tt ipole}\xspace}
\newcommand\ramiland{{\tt STIHOS}\xspace}
\newcommand\link{{\href{https://github.com/RamiAlBel/stihos}{GitHub-stihos}}\xspace}

\begin{document}

    \title{A deep learning algorithm for black hole spin estimation using hot-spot secondary images}
   
\titlerunning{Lensing of hot spots around black holes II}

\authorrunning{Yfantis et al.}

    \author{A. I. Yfantis 
          \inst{1}{*}
          \and
          R. Al-Belmpeisi \inst{2}*
    }

    \institute{Instituto de Astrofísica de Andalucía-CSIC, Glorieta de la Astronomía s/n, E-18008 Granada, Spain\label{inst1} \\
    \email{ayfantis@iaa.csic.es}
    \and Department of Applied Mathematics and Computer Science, Technical University of Denmark, Denmark \label{2}\\
    \email{ralbe@dtu.dk}\\
    {*}Authors have contributed equally in this work.}

   \date{Received: August 2026}

  \abstract
  {Sagittarius~A* exhibits frequent flaring activity across the electromagnetic spectrum that is often associated with a localized region of strong emission known as a hot spot.} 
   {We aim to train a deep learning model to provide a link between key parameters of this phenomenon -- hot-spot emission radius, and black hole inclination and spin -- to the observed angle difference between the primary and secondary image ($\Delta PA$) that present and future interferometric arrays could resolve.}
 {Using the general relativistic radiative transfer code \ipole, we generated a library of $\sim100.000$ models with varying system parameters and computed the position angle difference on the image plane between the primary and secondary images of the hot spot. 
We explore equatorial and non-equatorial circular orbits and evaluate our models against approximate observational constraints, including partial-orbit visibility and observational errors.}
{Our algorithm \ramiland shows remarkable accuracy in calculating spin and inclination from the majority of the observational tests we perform ($\sigma_{\alpha_*}=0.04,\,\sigma_i=2^{\circ}$), even in extreme conditions where only half of the orbit is visible. The off-equatorial estimation provides softer constraints in the absence of prior information. }
{ Our results demonstrate the importance of hot-spot observations for spacetime estimations. Given the increasing efforts to detect the photon-ring, our framework could prove valuable in interpreting the first observations of lensed emission.}


   \maketitle

\section{Introduction}

The supermassive black hole at the center of our Galaxy, Sagittarius~A* (\sgra), provides an excellent laboratory for studying accretion and general relativity. This is largely due to the Event Horizon Telescope (EHT), which has produced the first images of its shadow in total intensity \citep{eht:2022_paperI,eht:2022_paperII,eht:2022_paperIII,eht:2022_paperIV,eht:2022_paperV,eht:2022_paperVI} and in linear polarization \citep{eht_2024a,eht:2024b}, offering the closest view to date of the near-horizon region. In addition, the GRAVITY collaboration has provided the most precise constraints on the mass and distance of \sgra, measuring $M = 4.3 \times 10^6, M_{\odot} \pm 0.25\%$ \citep{Gravity2022} and $R_0 = 8.275,\mathrm{kpc} \pm 0.41\%$ \citep{gravity2021_sgraR}. Together, these measurements enable precision studies of the spacetime in the immediate vicinity of the black hole.

As discussed in the preceding paper \citealt{Y26} (hereafter \citetalias{Y26}), isolating robust spacetime signatures in black hole images requires mitigating contamination from plasma-dependent emission. In particular, finite resolution, interstellar scattering, and rapid variability of Sgr~A* complicate the disentanglement of spacetime effects. 

A typical example of this black hole spacetime signature is the photon ring, arising from photons on unstable bound orbits \citep{Bardeen_BH_1972}. In Kerr spacetime, the structure is determined solely by the black hole mass and spin, making it a promising probe for precision tests of gravity, but its observational separation from direct emission remains challenging. This feature has been extensively studied both theoretically \citep{Claude2001,JP2010,Gralla2019,Gralla2020_lensing,Prashant2024,Walia25} and observationally \citep{Johnson2020,Gralla2020_grtest,Wong2021,2025PhRvD.111f4075Z,2025PhRvD.112h3024Z}, and is a primary science target of the Black Hole Explorer (BHEX) mission \citep{Johnson_2024,BHEX2024a,BHEX2024b}. 

Following the logic in \citetalias{Y26}, we focus on a specific case of lensed emission, arising from secondary images of localized emission, often referred to as hot spots, instead of photon rings that encapsulate the whole disk. In this scenario, the secondary image remains localized, forming a crescent instead of a ring. This distinction between direct and lensed photons bypasses the shortcomings of disentanglement, enabling the use of null geodesic theory to extract information about the spacetime, and subsequently the dimensionless spin parameter, $\alpha_*=Jc/GM^2$. In particular, our study proposes a methodology that uses the angle difference $(\Delta PA)$ between the primary and secondary images to separate the spin effect from the orbital effect.  

On the motivation of localized emission, it is widely accepted that \sgra flares on a roughly daily basis, with some of these events resonating across the electromagnetic spectrum, exceeding the emission of the quiescent state by more than two orders of magnitude {\citep{gravity:2020c,sgra_xflares2019}}. Furthermore, the brightness centroid has appeared to orbit at some of these events in near-infrared observations (NIR), and the linear polarization ticks to rotate in a characteristic way of orbiting features in the flow \citep{gravity:2018,G23,Wielgus2022_LC,W22,eht:2022_paperII}. Numerous works connected these observational features with the hot-spot model, and the idea of a localized strong emission region orbiting around the black hole \citep[e.g.,][]{Brod-Loeb2006, Trippe2007, Hamaus2009,Tiede2020,Gelles:2021,Vos:2022,Vincent2023, gravityMichi, gravityAlejandra,ball:2021,aimar_2023_plasmoid,yfantis24a,yfantis24b,Antonopoulou24}, see also \cite{hotspot_rev26} for an extensive review.

Physical motivations for this hot spot phenomenon have been discussed in \citetalias{Y26}, along with the mathematical prior connecting $\dpa$ with $\alpha_*$, system inclination ($i$), and orbital velocity $\kc$. The connection to the emission radius $\rs$ was first expressed there, in the form of an empirical relation that relates $\dpa$ to all these parameters in a simple manner, as 

\begin{equation}
    \Delta PA \, [{}^\circ] = 180^\circ + \alpha_*(40-r_{\textrm{hs}}) - \left(\frac{1408-233\alpha_*}{r_{\textrm{hs}}^2} +35\right) \kc \,.
\label{eq:empirical}
\end{equation}

In what follows, we perform a systematic study of hot-spot models, enabled by our introduced Deep Learning (DL) framework \ramiland (Space-Time Inference via Hotspot Observations)\footnote{From the Greek stihos ($\sigma\tau\grave\iota \chi o \varsigma$), meaning verse or lyric.} and radiative transfer code \ipole, with the goal of obtaining robust and accurate estimates for $\alpha_*$ and $i$, while lifting the most notable limitations of \citetalias{Y26}, namely the averaging of $\dpa$ values for a full period and the equatorial orbits. Additionally, we examine partial hot-spot orbits as a proxy for realistic observations, where bright features might not be visible at all times. This is part of an emerging trend in accretion studies to use highly non-linear DL models to infer Sgr A* parameters, leveraging their emerging inference capabilities. Some examples of this include \cite{NN_saraetosi25,NN_janssen25,NN_feng26}.

In section \ref{sec:methods}, we discuss the computational methods used in this work, including a thorough description of our dataset containing hot spot models. Next, in Section \ref{sec:results}, our main findings on inference capabilities through our pipeline. Finally, in Sections \ref{sec:discussion}, \ref{sec:conclusion}, we discuss our findings and conclude.

\section{Methods}
\label{sec:methods}

In this section, we briefly describe the numerical setup for simulating hot-spot models. Then, we outline the architecture of the DL pipeline \ramiland, which uses hot-spot models to assess the limits of black hole parameter inference from $\dpa$ observations across the EHT, ngEHT, and BHEX eras. 

\subsection{Hot-spot simulations}

Hot-spot modeling is performed within the general-relativistic radiative transfer code \ipole \citep{Monika2018}. The hot spots produce purely synchrotron emission from a thermal electron population and are defined by fixed plasma parameters $n_e$, $\theta_e$, and $B$ (electron number density, temperature, and magnetic field strength), as we are interested only in geometric features. Their radius is set to $1.5M$ with a strict cut-off. They travel in circular orbits that lie in any plane parallel to the equatorial plane, for non-equatorial angular offsets with $\theta_z \in [-45^\circ, 45^\circ]$. This is motivated by works as \cite{aimar_2023_plasmoid, antonop025_spiral} where such trajectories arise from magnetic flux ropes, rotating in the equatorial plane, and extending along the jet sheath. The remaining free parameters in our models include the black hole spin $\alpha_*$, the hot-spot center $\rs$, the orbital velocity via the parameter $\kc$, capturing orbital velocities in the sub- and super-Keplerian regimes, and the system's inclination angle $i$.

In Table \ref{tab:params}, we list all parameters, including their ranges and cadence. We allowed combinations of values that produce periods from $P_{\text{min}}=20$ to $P_{\text{max}}=100$ min, since all observations have shown periods from $50$ to $90$ min \citep{W22,G23}. Additionally, shorter periods can exceed the speed of light near the horizon, and longer periods increase simulation time significantly. The method is expected to scale naturally to longer periods. Since all combinations amount to roughly 500 million models, we randomly sample across the parameter space, resulting in about 100,000 models. 
This avoids a fixed grid that performs worse in values outside of it. To ensure sufficient sampling we tested 3 subsets of the original library, and prediction error varies by a maximum of $20\%$, mostly attributed to the smaller size.

\begin{figure}[h!]
    \centering
        \includegraphics[width=0.901\linewidth,trim={0.2cm 0.0cm 0 0.2cm},clip]{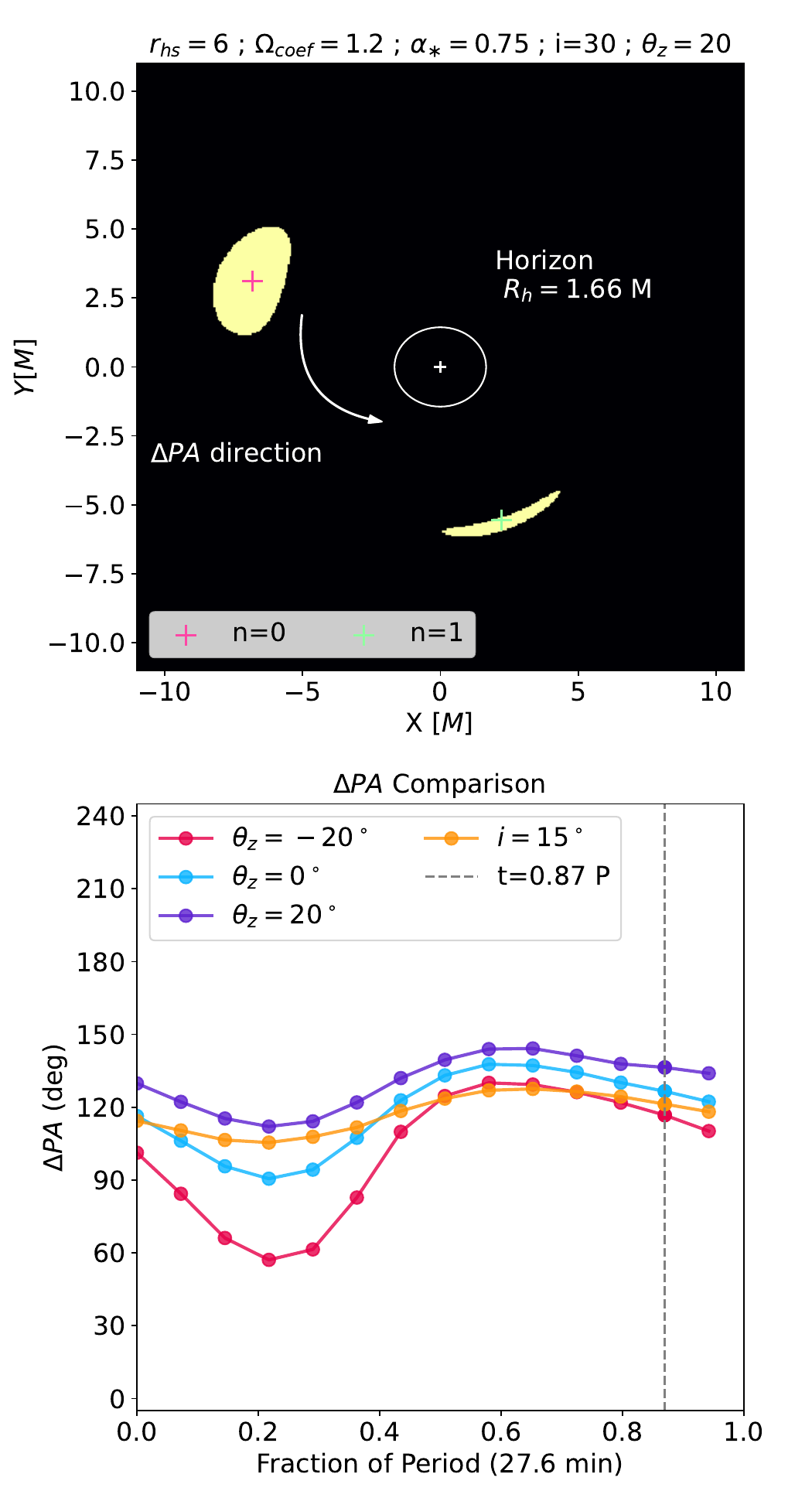}
   
    \caption{Top: Snapshot from a hot spot movie, with resolution 256x256 (as in the entire library), where $\Delta PA$ is calculated from the primary to the secondary counterclockwise. The setup parameters are visible in the title. Bottom: Curve of $\Delta PA$ measurement with respect to time for a full period for three models with the same azimuthal displacement $\theta_z$. The gray vertical line denotes the snapshot's time in the top panel.}
    
    \label{fig:ipole_image}
\end{figure}

The simulations are run with resolution $256\times256$, where we track the center of emission for $n=0$ and $n=1$ separately, as well as the $\Delta PA$ between them, as seen in the top panel of Figure \ref{fig:ipole_image}. Slow light is used in all simulations, a necessary step to account for the hot spot's motion as the photons travel to $n=1$.  

Each simulation is capped at a full orbital period, while all have a two-minute temporal cadence. 
An example of $\Delta PA$ curves is visible in the bottom panel of Figure \ref{fig:ipole_image}, using three models with different $\theta_z$ values ($-20^\circ,\,0,\,20^\circ$) and $i=30^\circ$, along with a $i=15^\circ$ model for comparison.  

\begin{table*}[h!]
  \centering
  \caption{Summary of randomly sampled parameter values in our simulated library.}
  \label{tab:params}
  \begin{tabular}{lccl}
  \hline\hline
  {Parameter} & {Range} & {Cadence} & {Description} \\
  \hline
  $\alpha_*$ & $(-1,\,1)$ & 0.01 & The spin of the black hole (negative values correspond to counter-rotating accretion) \\
  $\rs \, [M]$ & $[4,\,12]$ & 0.1 & Cylindrical radius of the hot spot center (projected onto the equatorial plane) \\
  $\kc$ & $[0.2,\,1.5]$ & 0.1 & Keplerian coefficient for orbital velocity ($1$ corresponds to Keplerian motion) \\
  $i\,[^\circ]$ & $[0,\,60]$ & 0.1 & Inclination angle of the system (0 is face-on) with respect to the observer \\
  $\theta_{z}\,[^\circ]$ & $[-45,\,45]$ & 1 & Angular offset of the hot spot from the equatorial plane (0 is equatorial) \\
  \hline
  \end{tabular}
\end{table*}

\subsection{\ramiland}

Our DL framework, \ramiland, is constructed to solve the following inverse problem of hot-spot simulations by answering the question: given observables from a hot-spot movie ($r_s$, $\Delta \mathrm{PA}(t)$, and period $T$), how reliably can the underlying system parameters, such as $\alpha_*$, $i$, and $\theta_z$ be inferred? 
Our proposed solution for learning the inverse mapping from observables to system parameters relies on a supervised DL model trained on our dataset of hot-spot realizations. We use an $80/10/10\%$ split to partition our models into training, validation, and test subsets, respectively. The framework can be found fully trained at \link\footnote{\url{https://github.com/RamiAlBel/stihos}}, along with example datasets from our library for retraining and experimentation.

\begin{figure}[ht]
    \centering
    \includegraphics[width=\linewidth]{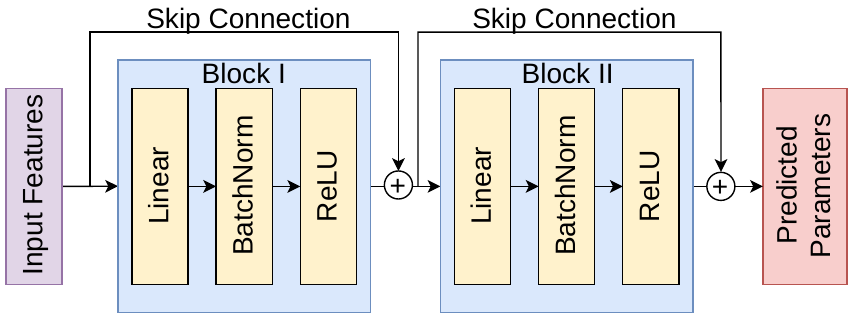}
    \caption{ The MLP architecture used across all experiments is displayed. Normalized input features propagate through two blocks; Each fully connected block consists of a fully connected layer consisting of 256 neurons, followed by batch normalization of intermediate features and a Rectified Linear Unit (ReLU) non-linear activation. Skip connections connect the input and output of fully connected blocks, adding the layer input to the block's output.
}
    \label{fig:model}
\end{figure}

\paragraph{Model architecture.}
We use a Multi-Layer Perceptron (MLP)-based architecture for parameter estimation across all experiments, as shown in Figure \ref{fig:model}. 
Our MLP model inputs consist of observables, including the hot-spot radius $\rs$, its orbital period $T$, and primary-secondary angles $\Delta PA (t_K)$ sampled on $t_K={N_{\Delta PA}T}/{10}$ for integer $N_{\Delta PA}\in[0,10)$.
Propagating inputs through the network yields the single-neuron output that estimates the target variables of black hole spin $\alpha_*$, or observer inclination $i$, under the assumption of equatorial orbits. If we further assume non-equatorial hot-spot orbits, we also provide predictions for the axial hot-spot distance $z$. 
Monte Carlo Dropout is used during training to enhance model generalization by randomly dropping 10\% of the network's neurons. Ensuring stable gradients during optimization requires scaling input and output features to have zero mean and unit variance.

We model \ramiland's function-approximation predictions with a shallow MLP for two main reasons: the highly non-linear nature of the $\dpa$-$\alpha_*$ dependence and the linear scaling of computational costs $\mathcal{O}(N)$ with the number of parameter-space samples.
While Gaussian processes are an appealing modeling alternative to shallow MLPs, their straightforward error computation and wide adoption in the community are offset by their $\mathcal{O}(N^3)$ scaling, which makes them suboptimal in our analysis and hints at the edge of DL-based modeling for future inference tasks. 

\paragraph{Optimization.}
The model parameters $\Theta$ are optimized to produce predictions that more closely match the ground-truth distribution during training. Therefore, the DL model used can be understood as a function $f$ that maps observables $x$ to predicted systemic parameters $\hat{y}$, given by:

\begin{equation}
\begin{aligned}
    \hat{y}= f(x;\Theta),
\end{aligned}
\end{equation}

Post-training, the optimal model weights $\Theta^*$ minimize the loss function (i.e., the penalty), and the model is evaluated on the independent validation set. In this framework, our loss is given by the mean squared error, given by:

\begin{equation}
\begin{aligned}
\mathcal{L}=L_1(y,\hat{y})=|y-\hat{y}|^2=|y-f_\Theta(x)|^2
\end{aligned}
\end{equation}

All networks are trained for up to 200 epochs using the Adam optimizer, starting with a learning rate of $10^{-3}$ and a weight decay of $10^{-4}$. We use a 10-epoch patience for halving the learning rate and a 20-epoch limit for early stopping.

\subsection{Observational noise and implementation}

Uncertainty computation for DL predictions is more challenging than standard statistical approaches, such as frequentist and Bayesian methods, due to the intrinsic non-linearity of activation functions in these models. Consequently, we explore two error-handling strategies, which are discussed below.

The first approach perturbs the input data during training by adding a predetermined noise to the inputs, modeled as a variance $\sigma_{x_i}$ for each input parameter {$x_i \in (\rs, T,  \dpa)$}.  The standard deviation of prediction errors $std(\hat{y}-y_{GT})$ under inputs $x_i$ with variance $\sigma_{x_i}$ can be treated as the uncertainty of the prediction under observational noise $\sigma_{x_i}$. 
The drawback of this method is the a priori choice of a fixed variance for the model inputs, which can be circumvented by training multiple models on a grid of input variances.
This grid can then be used to estimate prediction errors for any observational noise in the inputs $x_i$ during inference via trilinear interpolation across grid combinations, yielding estimates for configurations not calculated during training.
While the Gaussian noise injection is straightforward for $\rs$ and $T$, adding noise to $\Delta PA(t)$ should further respect temporal continuity. This motivates the use of a radial basis function (RBF) kernel to sample smooth noise from a Gaussian process. The RBF kernel correlates noise injection at two time-points $t_i$ and $t_j$ through:

\begin{equation}
    K(t_i,t_j)=\sigma^2 e^{-\frac{(t_i-t_j)^2}{2\tau^2}},
\end{equation}
where $\tau$ defines the correlation timescale, a parameter controlling the smoothness of the noisy $\Delta PA (t)$. Notice that $K(t_i,t_i)=\sigma^2$, suggesting that the noise added is equivalent to that of standard deviation $\sigma$. We use $\tau=0.3$ $T$ across all experiments. While the choice of kernel $\tau$ can shift prediction errors by roughly $\pm 70\%$, most modern imaging algorithms include some form of  regularization for EHT movies, motivating the kernel method for this test.

The second error-handling strategy utilizes DL models trained on noise-free inputs. During inference, prediction uncertainty is estimated using a Monte Carlo chain, with the observational errors as priors.
This strategy essentially performs multiple forward passes, each taking inputs modified by observation noise, and interprets the resulting variance as prediction uncertainty. A convergence analysis of up to 50,000 Monte Carlo steps reveals that 2,000 steps are sufficient for robust estimation of prediction uncertainty, without significant computational overhead.

This second approach concentrates the computational overhead of uncertainty estimation during inference rather than during training, allowing greater flexibility in combining uncertainties across observables. However, due to the small size of our models, inference is computationally cheap, enabling the effective and quick implementation of both uncertainty estimation strategies and making inference take just a few seconds, even on a single CPU.

To obtain an estimate of typical errors in Sgr A* observations, we consider the measurement error of the ring diameter $d= 51.8 \pm 2.3$ $\mu as$ \citep{eht:2022_paperI,eht:2022_paperIV}. If we consider two features, $f_0$ and $f_1$ (essentially the $n=0,1$ hot-spot images), with locations ($x_0,\,y_0$), ($x_1,\,y_1$) respectively, their relative angle $\Delta PA$, and its standard deviations are given as:

\begin{equation}
\label{eq:errors}
\Delta PA = \arctan\left( \frac{\Delta y}{\Delta x} \right), \qquad
\sigma_{\Delta PA} = \frac{2\sigma_{\Delta xy}}{L} \; \text{[rad]},
\end{equation}
where $\Delta y$ represents the difference $y_0-y_1$ (same for $\Delta x$), $L$ the length between the two features on the observer's screen, and $\sigma_{\Delta xy}$ the common positional uncertainty (in the case of $\sigma_{\Delta x}=\sigma_{\Delta y}$). Note that this error does not take into account any additional uncertainty from the imaging methods, thus in practice we also double $\sigma_{\Delta PA}$ in Sections \ref{sec:results_noise}, \ref{sec:Mock} to account for this to some extent.

\subsection{Experiments}

To assess the capabilities of the \ramiland framework for identifying key system parameters in supermassive black hole (SMBH) systems, we perform a series of experiments, grouped into five categories of increasing complexity, as shown in Table \ref{tab:experiments}.

\begin{table}[t]
\caption{A summary of the experimental setups.}
\label{tab:experiments}
\begin{tabular}{ccccc}
\hline
{Exp} & {Equatorial} & {Full orbit} & {Inputs} & {Predictions} \\
\hline
\noalign{\vspace{0.05cm}}
I   & $\checkmark$ & $\times$     & $r,\,T,\,\overline{\Delta\mathrm{PA}}$         & $\alpha_*$ \\
II  & $\checkmark$ & $\checkmark$ & $r,\,T,\,\Delta\mathrm{PA}(t)$        & $\alpha_*,\,i$ \\
III & $\checkmark$ & $\times$     & $r,\,T,\,\Delta\mathrm{PA}(t)$         & $\alpha_*,\,i$ \\
IV  & $\times$     & $\checkmark$ & $r,\,T,\,\Delta\mathrm{PA}(t)$        & $\alpha_*,\,i,\,\theta_z$ \\
V   & $\times$     & $\times$     & $r,\,T,\,\Delta\mathrm{PA}(t)$         & $\alpha_*,\,i,\,\theta_z$ \\
\hline
\end{tabular}
\tablefoot{
    The experiments performed (Exp) are categorized by the neural network input features (Inputs), the predicted parameters (Predictions), and the underlying assumptions about partial visibility (Full Orbit) and the azimuthal angle of hot-spot orbits relative to the disk plane (equatorial).
  }
\end{table}

Experiment I: “Orbit-aggregated $\dpa$'' aims to reproduce the results of \citepalias{Y26} using their intermediate result: that averaging all $\Delta PA$ values over a full period is equivalent to measuring $\Delta PA$ face-on ($i=0$). Additionally, as in all cases, the hot-spot's orbital radius $\rs$ and its period $T$ are required to estimate $\alpha_*$. This setting enables validation of our framework results and once again demonstrates the strength of this approach. 

Experiment II: “Equatorial Full Orbits” represents a significant extension. By incorporating the time-resolved positional angle difference between the hot spot and its secondary image, $\Delta \mathrm{PA}(t)$ (with the input given as a time series of $\Delta \mathrm{PA}$ values), we enable the simultaneous inference of $i$ and $\alpha_*$.

Experiment III: “Equatorial Partial” examines the framework's performance in scenarios where only a fraction of the full orbit is observed. This is achieved by inferring the target parameters from orbital segments spanning from $10\%$ to $100\%$ of the full trajectory. This setup is essential for assessing performance under more realistic observational constraints.

Experiments IV and V: “Non-Equatorial Orbits” generalize Experiments II and III, respectively, by considering hot-spot trajectories displaced from the equatorial plane. This setup not only enables the estimation of $\alpha_*$ and $i$ in a more general configuration, but also allows for the inference of $\theta_z$, a parameter with significant impact on the nature and behavior of hot spots.

\section{Results}
\label{sec:results}

To assess the framework's inference capabilities, our test metrics include the mean absolute error (MAE), the standard deviation ($\sigma$), and the coefficient of determination ($R^2$), all computed between the predicted and ground-truth values for each inferred parameter. 
In what follows, we first touch upon the epistemic uncertainty of our method, assessing our method in the idealized zero observational noise scenario, followed by an analysis of the effects of observational noise on model predictions, and finally, virtual observations showcasing expected constraints from our model under the presence of realistic observational errors expected from different arrays.

Experimental metrics are reported throughout as mean and standard deviation estimates, obtained by retraining the networks 3 times per experiment while varying stochastic components in data sampling and noise injection. 

\subsection{Parameter estimation with no errors}
\label{sec:results_no_noise}

\begin{figure}[ht!]
	\centering
	\includegraphics[width=\columnwidth]{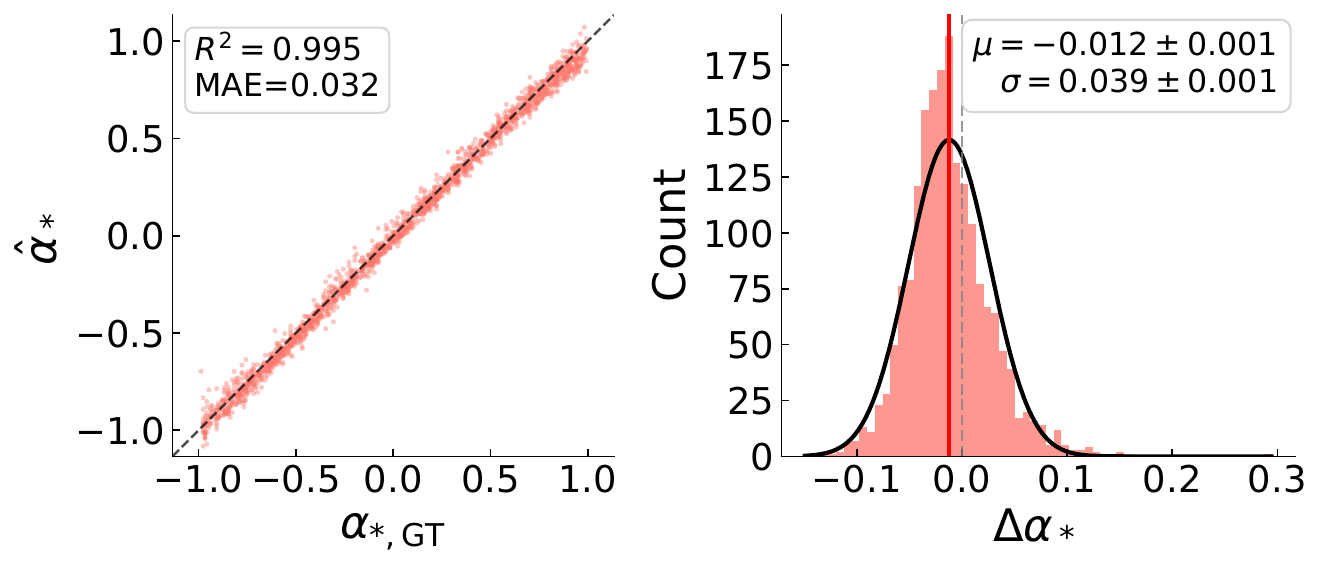}
	\hfill
	\caption{Experiment I SMBH spin predicted vs.\ actual values. Left: Prediction ($\hat{\alpha}_*$) vs ground truth ($\alpha_{*,\text{GT}}$).
		Right: Prediction error $\Delta \alpha_* = \hat{\alpha}_* - \alpha_{*,\text{GT}}$, as in all following Figures; $\mu=-0.01$ and $\sigma=0.04$ represent mean and standard deviation.} 
	\label{fig:exp1_pred_vs_actual}
\end{figure}

\paragraph{Experiment I.}
The results for the spin recovery from equatorial face-on orbit-averaged $\Delta PA$ are displayed in Figure \ref{fig:exp1_pred_vs_actual}. Please note that $\Delta$ refers to predicted minus ground truth for the remainder of this text. It is clear that $\alpha_*$ can be recovered near perfectly in the noiseless setting, with an MAE of $0.032$, $R^2=0.995$, and $\sigma_{\alpha_*}=0.04$. This once again verifies the method of averaging $\Delta PA$ values from a full period to approach $\Delta PA$ in the $i=0$ case.

\begin{figure}[ht!]
	\centering
	\includegraphics[width=\columnwidth]{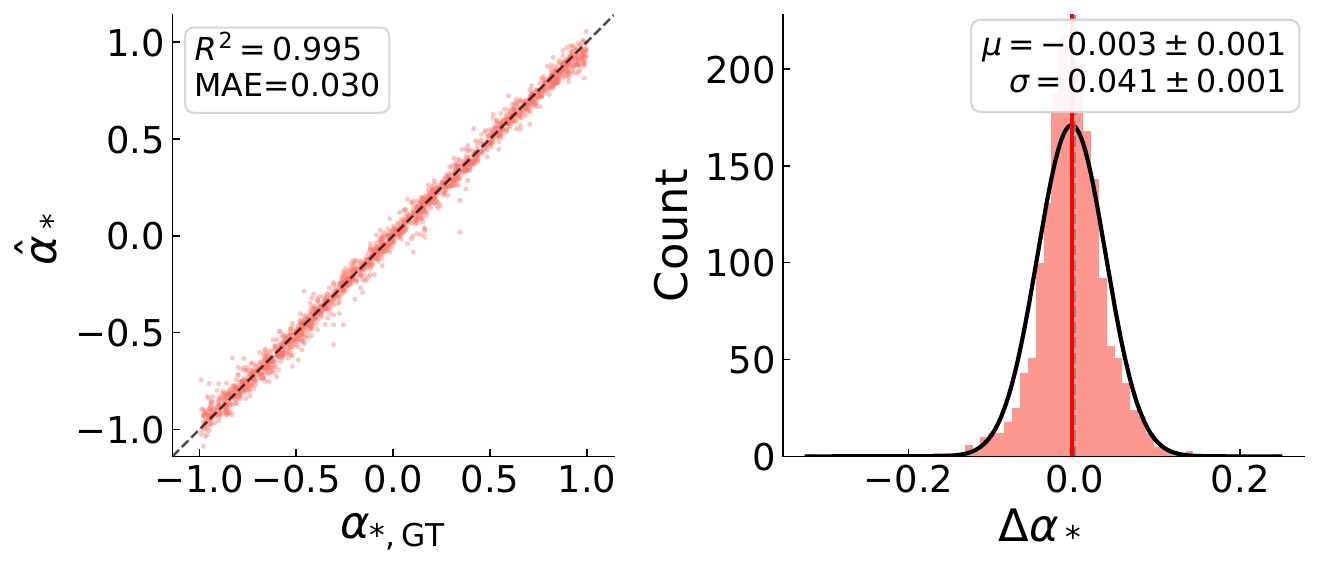}
	\includegraphics[width=\columnwidth]{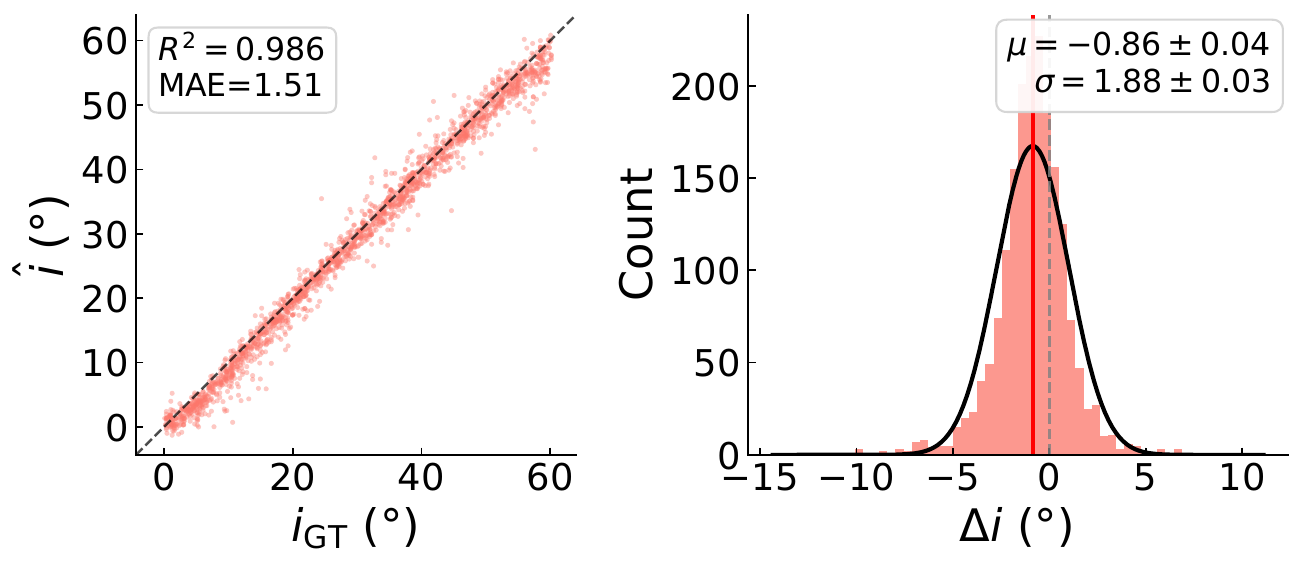}
	\includegraphics[width=0.881\columnwidth]{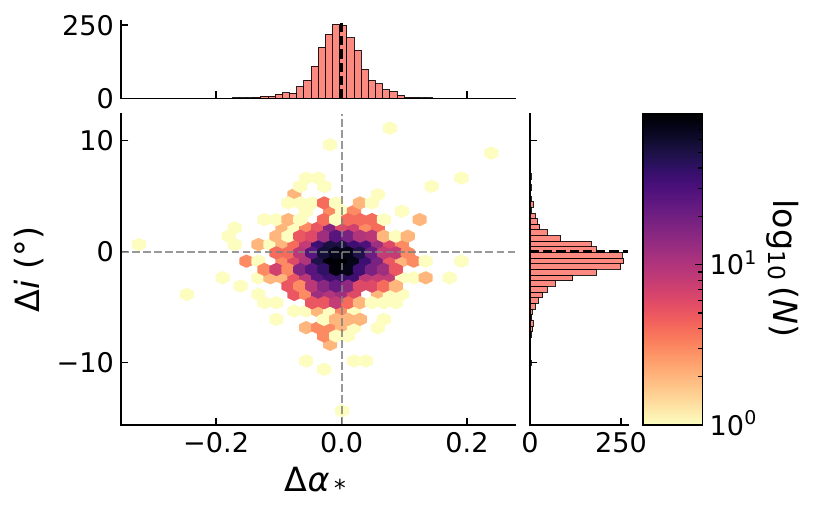}
	
	\caption{Experiment II predictions vs.\ ground truth.
		Top plots similar to Figure \ref{fig:exp1_pred_vs_actual} for $\alpha_*$ and $i$. Bottom joint distribution plot for possible trends.}
	\label{fig:exp2_pred_vs_actual}
\end{figure}

\paragraph{Experiment II.}
The results for inferring $i$, $\alpha_*$ utilizing full orbit time-curves of $\Delta PA$ (as can be seen in Fig. \ref{fig:ipole_image}), are visible in Fig. \ref{fig:exp2_pred_vs_actual}. The accuracy of $\sigma_{\alpha_*}=0.04$ remains identical to Exp. I, while errors on inclination predictions display errors of $\sigma_{i}=1.9^\circ$. Interestingly, $\mu_i = -0.9^\circ$, reflecting a small under-prediction from \ramiland in inclination estimations. This experiment shows, for the first time, the use of $\Delta PA$ curves to obtain highly accurate estimates of inclination and spin across virtually the entire parameter space for equatorial, circular hot spots.

\begin{figure}[ht!]
	\centering
	\includegraphics[width=0.913\columnwidth]{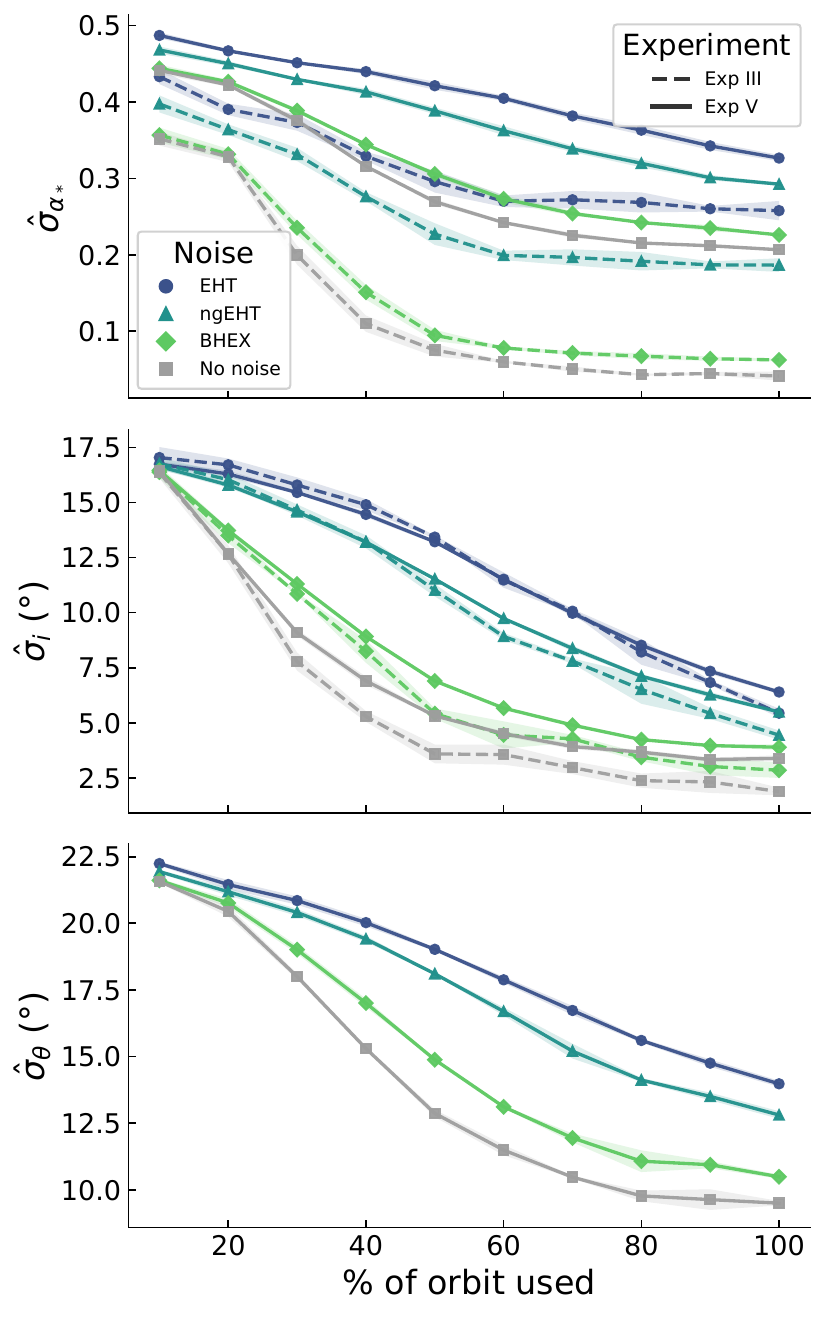}
	\caption{Experiment III \& V: error standard deviation $\sigma$ as a function of orbital
		coverage fraction for parameters $\alpha_*$, $i$, and $\theta_z$ (Exp V) case. Dashed lines represent the noiseless control, while solid ones represent the noisy setting.}
	\label{fig:exp35_sweep}
\end{figure}

\paragraph{Experiment III.}
SMBH observations with resolved hot-spot features are not expected to necessarily contain a full orbital period. For example, in radio, the disk might obscure the secondary image in certain parts of the orbit. It is therefore important to assess how inference performance declines as a function of the orbital coverage fraction, assuming equatorial orbits. In Figure \ref{fig:exp35_sweep}, we show the performance of this analysis with dashed gray lines on the top two panels. The expected trend of increasing errors with decreasing orbital coverage fraction is reflected in the results, as observing $10\%$ of the full orbit can lead to errors about $100$ times higher than in the $100\%$ case.
What is unexpected and advantageous for future observations is that the inferring power remains high for fractions down to $50\%$ of a full orbit (with $\sigma_{\alpha_*}=0.08$, $\sigma_{i}=4$).

\begin{figure}[ht!]
	\centering
	\includegraphics[width=0.86\columnwidth]{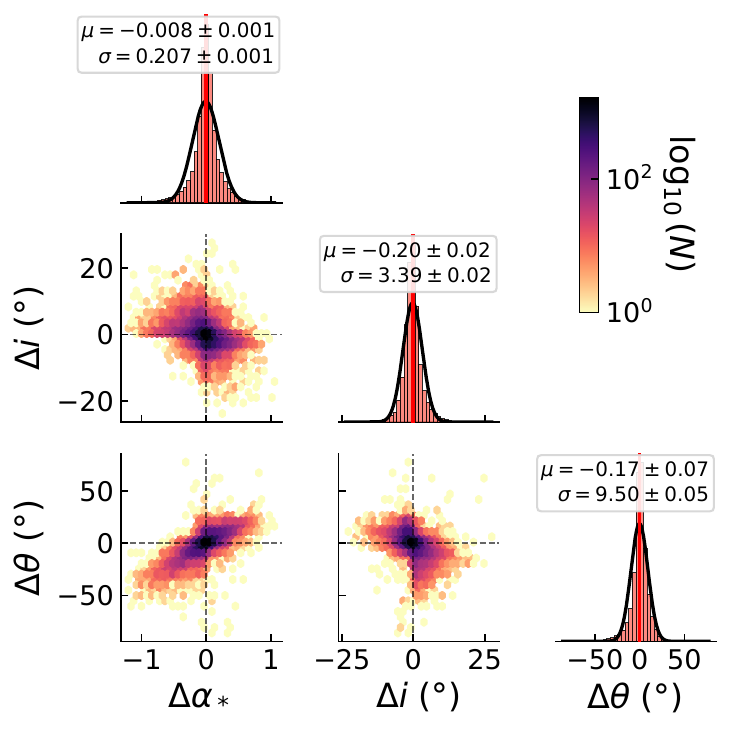}\\
	\caption{Experiment IV predictions vs.\ ground truth. Triangle plot, similar to Figure \ref{fig:exp2_pred_vs_actual}, for $\alpha_*$, $i$, and $\theta_z$. }
	\label{fig:exp4_pred_vs_actual_clean}
\end{figure}

\paragraph{Experiment IV.}
The near-excellent predictive performance of our framework so far has relied upon the assumption of equatorial orbits. Relaxing this constraint to include non-equatorial orbits expands the parameter space, which in turn leads to degeneracies that complicate the inference of spin and inclination. A characteristic example in the bottom panel of Figure \ref{fig:ipole_image} illustrates how off-equatorial orbits can mimic higher inclination. 
This experiment also provides predictions for the out-of-plane angular offset $\theta_z$, the results of which are jointly summarized in Figure \ref{fig:exp4_pred_vs_actual_clean}, where a corner plot provides the distribution of joint errors across predicted parameters. The assumption of non-equatorial orbits mostly affects the inference of $\alpha_*$, as
prediction errors $\sigma_{\alpha_*}$ rise from 0.04 to 0.22, while the difference is smaller for $i$, with $\sigma_{i}$ errors rising from $1.9^\circ$ to $2.87^\circ$. The errors on $\theta_z$ are quite large at $\sigma_{\theta_z}= 9.55$, providing only soft constraints on the location with respect to the accretion flow, for example, on the jet-sheath or close to the equator.

\paragraph{Experiment V.}
The combination of non-equatorial geometry with partial visibility yields the most general setting. Its results are summarized jointly with Exp. III in Fig. \ref{fig:exp35_sweep}. 
The parameters exhibit similar trends to those in Exp. III, but $\sigma_{i}$ is less sensitive to the hot-spot's off-equatorial assumption, compared to $\sigma_{\alpha_*}$. Additionally, the plateau for $\theta_z$ is smaller, reaching only $70\%$ of orbits.

\subsection{Uncertainty under observational noise}
\label{sec:results_noise}

In this section, we assess the aleatoric uncertainty in \ramiland's model predictions on local spacetime properties, meaning performance under noisy data. To comparatively perform this test across experiments, we turn to available missions for possible observations, including the EHT, as well as plans for the ngEHT and BHEX, for a hot spot at $\rs\sim5$ M.

First, in the case of the current EHT array, an image centroid accuracy following the ring diameter measurement of $d= 51.8 \pm 2.3\, \mu as$ \citep{eht:2022_paperI,eht:2022_paperIV} can be expected, where ${1M = 5.227 \, \mu\text{as}}$. This suggests a centroid error of $0.5M$ for $\rs$, which after doubling to adopt a more conservative estimate, yields $\sigma_{\rs} = 1M$ and correspondingly $\sigma_{\Delta PA} = 13.5^\circ$ (using equation \ref{eq:errors}). The current temporal resolution that dictates the uncertainty in the period is set by the ALMA light curve \citep[see][]{Wielgus2022_LC}, with $\sigma_P \sim 1$ min.

Moving on to the case of the next-generation EHT array (ngEHT), we can expect an improvement of a factor of 1.5 in terms of angular resolution \citep{ngeth23}, which translates to a linear drop in image measurement errors by a factor of roughly one-third, compared to the current EHT array. This suggests that $\sigma_{\rs}  \sim 0.65M$ and therefore $\sigma_{\dpa}\sim 9 ^\circ$. As ngEHT also includes ALMA, we again assume $\sigma_P\sim 1$ min.

Lastly, we consider the observational limits of the proposed BHEX mission \citep{Johnson_2024}, which offers significant jumps in expected resolution. Following the reported values from \cite{bhex24_pring}, the photon ring is expected to be measured with less than $r_{\text{ph}} \pm 0.1$ accuracy, and with a constant cadence of 10 seconds.
Our estimates for BHEX therefore translate to $\sigma_{\rs}\sim0.1 M$, $\sigma_{\dpa} \sim 1.5 ^\circ$, and $\sigma_P \sim 0.5$ min.

In Table \ref{tab:results_summary}, we present the results for each of these arrays and their respective errors. Note that in different $\rs$ and $\Delta PA$ values, the errors would change, as shown in equation \ref{eq:errors}, but we use indicative errors for $\rs\sim5$ for all tests. For conciseness, we present results only from BHEX, except in Figure \ref{fig:exp35_sweep}, where we show all arrays. The $\sigma$ values at $100\%$ are effectively the errors for Exp. II and IV, where the expansion of uncertainty in inference for increasing noise is visualized.

\begin{table*}[ht!]
    \centering
    \caption{ Performance of experiments I, II, and IV for different noise budgets.}
    \label{tab:results_summary}
    \resizebox{0.9\textwidth}{!}{
    \setlength{\tabcolsep}{2.5pt}
    \begin{tabular}{ll cccc cccc cccc cccc}
        \hline\hline
        \rule{0pt}{2.5ex}
        & & \multicolumn{4}{c}{{No noise}} & \multicolumn{4}{c}{{BHEX}}
          & \multicolumn{4}{c}{{ngEHT}} & \multicolumn{4}{c}{{EHT}} \\
        & & \multicolumn{4}{c}{$(0,\,0,\,0)$} & \multicolumn{4}{c}{$(0.1,\,2,\,0.5)$}
          & \multicolumn{4}{c}{$(0.65,\,9,\,1)$} & \multicolumn{4}{c}{$(1,\,14,\,1)$} \\
        \cline{3-6}\cline{7-10}\cline{11-14}\cline{15-18}
        \rule{0pt}{2.5ex}
        {Exp} & {Target}
          & $\mu$ & $\sigma$ & MAE & $R^2$
          & $\mu$ & $\sigma$ & MAE & $R^2$
          & $\mu$ & $\sigma$ & MAE & $R^2$
          & $\mu$ & $\sigma$ & MAE & $R^2$ \\
        \hline
        \rule{0pt}{2.5ex}
        I & $\alpha_*$ & $-0.012$ & $0.039$ & $0.032$ & $0.995$ & $-0.009$ & $0.075$ & $0.060$ & $0.983$ & $-0.010$ & $0.252$ & $0.201$ & $0.806$ & $-0.019$ & $0.348$ & $0.278$ & $0.630$ \\
        \hline
        \rule{0pt}{2.5ex}
        II & $\alpha_*$
          & $-0.003$ & $0.041$ & $0.030$ & $0.995$
          & $-0.006$ & $0.062$ & $0.049$ & $0.988$
          & $-0.004$ & $0.186$ & $0.147$ & $0.894$
          & $+0.001$ & $0.258$ & $0.202$ & $0.796$ \\
           & $i$ [$^\circ$]
          & $-0.86$ & $1.88$ & $1.51$ & $0.986$
          & $-0.56$ & $2.78$ & $2.12$ & $0.972$
          & $-0.57$ & $4.40$ & $3.31$ & $0.934$
          & $-0.78$ & $5.44$ & $4.08$ & $0.899$ \\
        \hline
        \rule{0pt}{2.5ex}
        IV & $\alpha_*$
          & $-0.008$ & $0.207$ & $0.138$ & $0.868$
          & $-0.001$ & $0.226$ & $0.158$ & $0.843$
          & $-0.003$ & $0.292$ & $0.215$ & $0.737$
          & $-0.008$ & $0.327$ & $0.245$ & $0.671$ \\
           & $i$ [$^\circ$]
          & $-0.20$ & $3.39$ & $2.33$ & $0.961$
          & $-0.11$ & $3.89$ & $2.77$ & $0.949$
          & $-0.27$ & $5.49$ & $4.17$ & $0.898$
          & $-0.16$ & $6.40$ & $4.90$ & $0.862$ \\
           & $\theta$ [$^\circ$]
          & $-0.17$ & $9.50$ & $6.31$ & $0.854$
          & $-0.17$ & $10.49$ & $7.13$ & $0.823$
          & $-0.39$ & $12.81$ & $9.29$ & $0.735$
          & $+0.24$ & $13.98$ & $10.31$ & $0.685$ \\
        \hline
    \end{tabular}
    }
    \tablefoot{Bias ($\mu$), scatter ($\sigma$), mean absolute error (MAE), and $R^2$ for the BHEX, ngEHT, and EHT instrument error budgets, with the corresponding noise included during training. Noise budgets are quoted as $(\sigma_{r}\,[M],\ \sigma_{\Delta\mathrm{PA}}\,[^\circ],\ \sigma_{T}\,[\mathrm{min}])$.}
\end{table*}

\begin{figure}[ht!]
	\centering
	\includegraphics[width=\columnwidth]{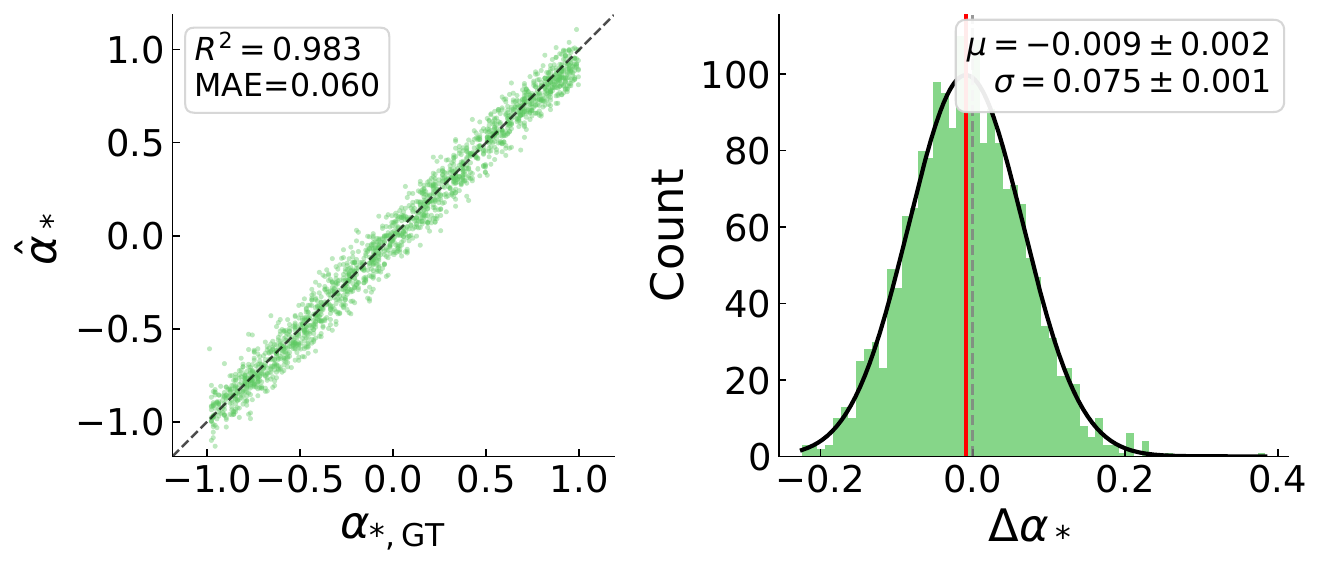}
	\hfill
	\caption{Experiment I spin prediction using errors during training.}
	\label{fig:exp1_noise}
\end{figure}

\begin{figure}[ht!]
	\centering
	\includegraphics[width=\columnwidth]{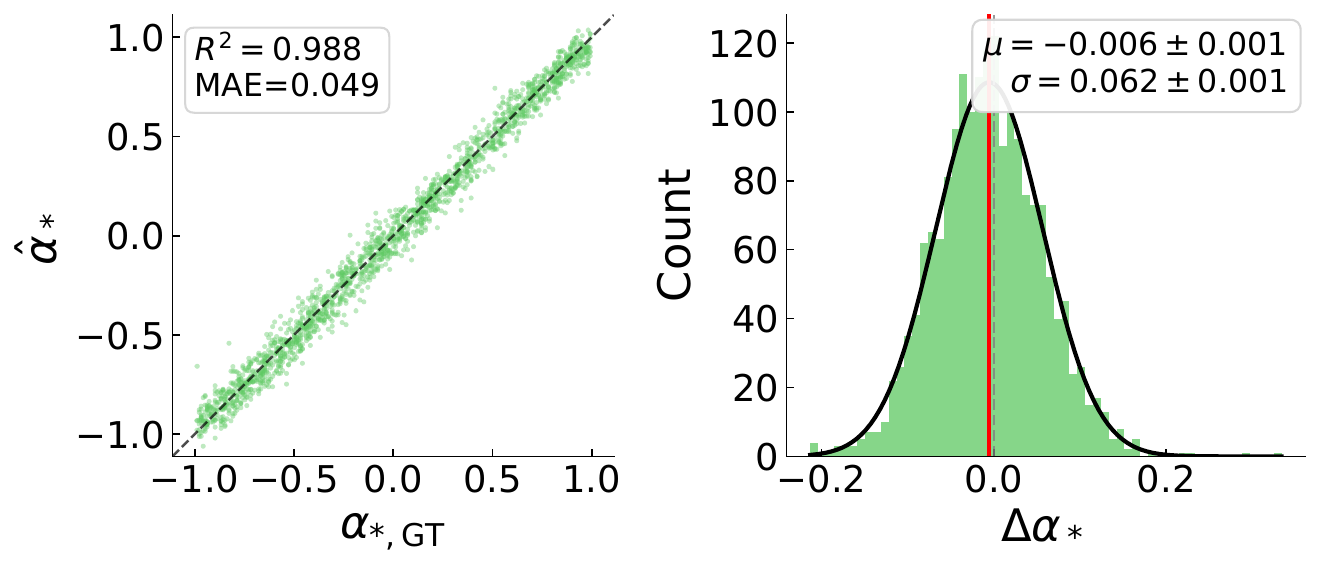}\\[4pt]
	\includegraphics[width=\columnwidth]{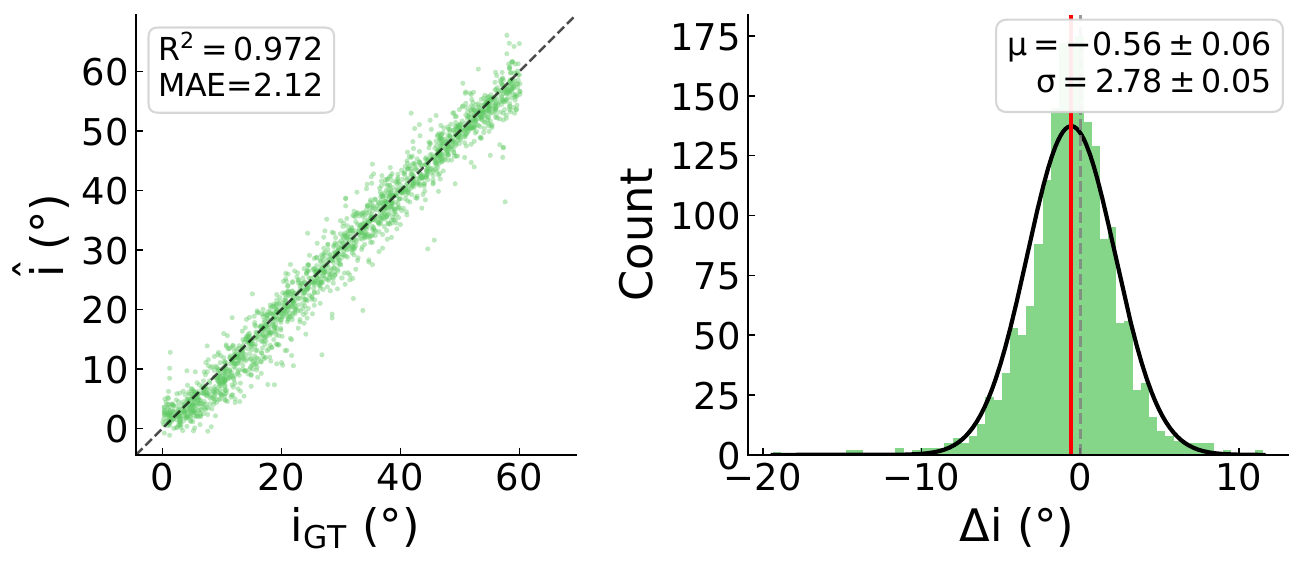}
	\includegraphics[width=0.891\columnwidth]{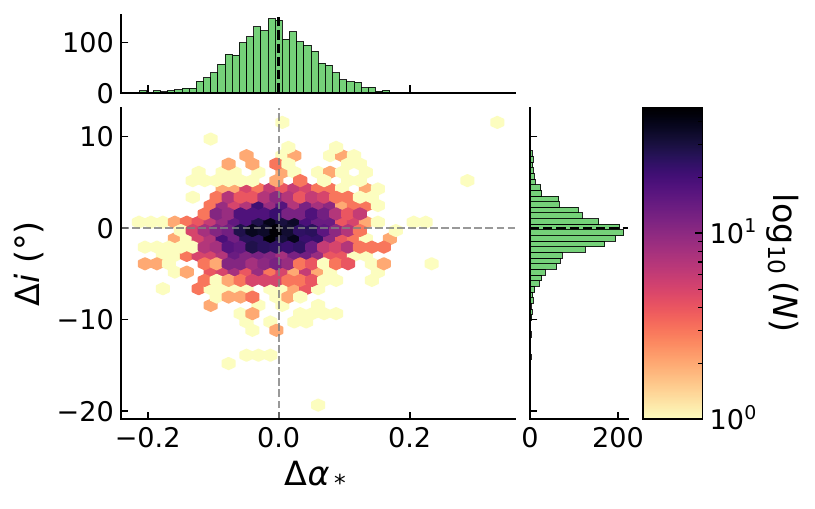}
	
	\caption{Experiment II predictions vs.\ ground truth using errors during training. Top: black hole spin parameter $\alpha_*$ (MAE $= 0.049$, $R^2 = 0.988$).
		Middle: inclination $i$ (MAE $= 2.12^\circ$, $R^2 = 0.972$). Bottom: joint distribution of the errors.}
	\label{fig:exp2_noisy}
\end{figure}

\begin{figure}[ht!]
	\centering
	\includegraphics[width=0.86\columnwidth]{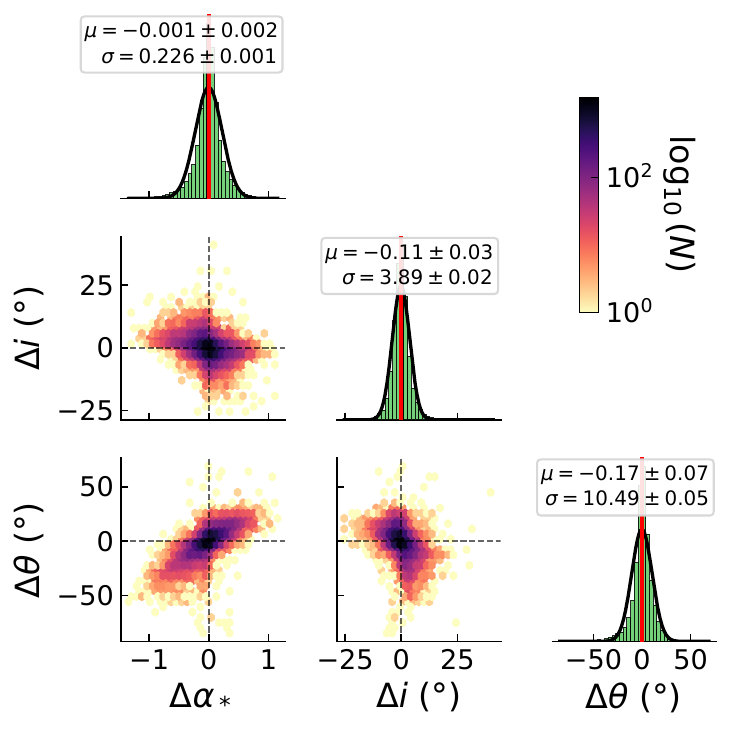}\\
	\caption{Experiment IV predictions vs.\ ground truth. Triangle plot, similar to Figure \ref{fig:exp4_pred_vs_actual_clean}, but using noise in training.}
	\label{fig:exp4_pred_vs_actual_noisy}
\end{figure}

The introduction of BHEX noise for Exp. I,II, and IV are visible in Fig. \ref{fig:exp1_noise}, \ref{fig:exp2_noisy} and \ref{fig:exp4_pred_vs_actual_noisy} respectively.
Under the assumption of equatorial hot-spot orbits, the extraction of SMBH spin $\alpha_*$ and $i$ remains highly feasible since $\sigma_{\alpha_*}$ increases two times from $0.03$ to $0.07$ and $\sigma_i$ even less from two to three degrees.

When assuming non-equatorial hot-spot orbits, the growth is similar in absolute terms, as $\alpha_*$ predictions increase again by $ \Delta \sigma_{\alpha_*} \sim 0.04$, and inclination by $ \Delta \sigma_{i} \sim 1$, translating to a smaller relative difference. While for $\sigma_{\theta_z}\sim 10^\circ$, the trend is similar, with a mild increase of two degrees. 
For a comprehensive summary of model performance degradation under all levels of observational noise across Exp. I, II, and IV (see Table \ref{tab:results_summary}). There, it is evident that the effects of larger observation uncertainties also significantly push the inference to $\sigma_{\alpha_*}\sim 0.3$ and $\sigma_i\sim 5^\circ$.

In the cases of Exp. III and V, the BHEX noise effects are visible in the green curves of Fig. \ref{fig:exp35_sweep}, where once again they closely match the errorless case, with the flat increase persistent in all period fractions down to $30\%$, where below that fraction uncertainties from the curves are matching the observational ones, hence $\sigma_i$ values converge.

\subsection{Virtual observations}
\label{sec:Mock}

We further evaluate \ramiland against two virtual observations, with $\rs \sim 5$ and $\Delta PA \sim 170^\circ$ and $ 90^\circ$ for positive and negative spins, respectively, using the Monte Carlo method for error estimation. 
Following the same logic as in \ref{sec:results_noise}, a summary of both virtual observations and their expected errors per observational array is provided in Table \ref{tab:mocks_errors}.

\begin{table}[h]
	\centering
	\caption{Virtual observations and assigned uncertainties.}
	\label{tab:mocks_errors}
	\begin{tabular}{lccc}
		\toprule
		Instrument & $\sigma_{\rs}\,[M]$ & $\sigma_{\Delta PA}\,[^\circ]$ & $\sigma_P\,[\mathrm{min}]$ \\
		\midrule
		EHT     & 1.0  & 14   & 1   \\
		ngEHT   & 0.65 & 9    & 1   \\
		BHEX    & 0.1  & 2  & 0.5 \\
		\midrule
		Virtual data & $r_s\,[M]$ & $\Delta PA\,[^\circ]$ & $P\,[\mathrm{min}]$ \\
		\midrule
		1 & 6.0 & $\sim 170$ & 55 \\
		2 & 4.5 & $\sim 90$  & 30 \\
		\bottomrule
	\end{tabular}
	\tablefoot{$\Delta PA$ values vary significantly for different $i$ and $\theta_z$, so the reported value is indicative for the $i=0$ case.}
\end{table}

\begin{figure*}[h!]
	\centering
	\includegraphics[width=0.95\linewidth]{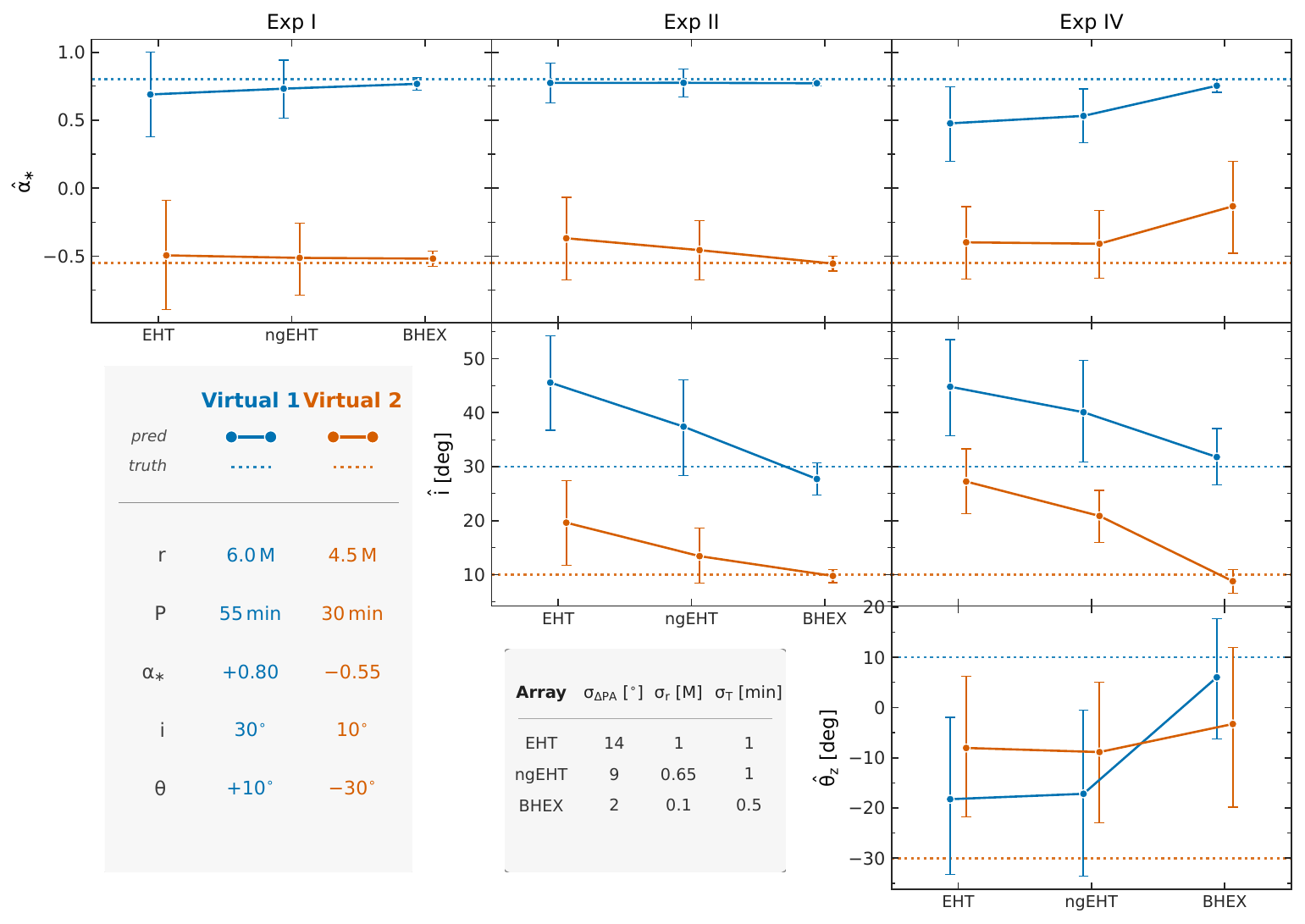}
	\caption{From left to right, results for Exp. I, II, and IV. From top to bottom, the panels correspond to $a_*$, $i$, and $\theta_z$. The y-axis displays estimated values and $1\sigma$ uncertainties, where dotted lines denote the true value for each virtual observation. The x-axis shows the three instruments. All virtual observation parameters are displayed in the bottom-left corner, and instrumental errors are in the bottom-right table.}
	\label{fig:mock_res}
\end{figure*}

The results of this analysis are reported in Table \ref{tab:mocks_tier_grid_3x3_absolute_2sigma} and visualized better in Figure \ref{fig:mock_res}, with blue and orange denoting virtual 1 and virtual 2, respectively. From left to right, we present results for Exp. I, II, and IV. From top to bottom, the panels correspond to $\alpha_*$, $i$, and $\theta_z$. The y-axis displays estimated values and $1\,\sigma$ confidence, where dotted lines denote the true value for each virtual observation. The x-axis shows the three instruments. All virtual observation parameters are displayed in the bottom left corner.

The error bars improve progressively for each array, with smaller errors, except for Exp. IV. This is due to the strong degeneracy between these three parameters $\alpha_*$, $i$, and $\theta_z$, as we discuss further in the next section. Despite that, in virtual 1 there is still convergence but it seems parameter-set dependent. In both virtual 1 and 2, $\sigma_{\alpha_*}$ starts at $\sim0.3$ for EHT and reaches $\sigma_{\alpha_*}\leq0.05$ for BHEX, with the median value also converging for each array. Inclination estimation starts farther away for EHT, while still covering the truth with $2\sigma$ confidence, and converges closer with each array, reaching $\sigma_i\leq2$. The confidence levels are comparable to those of the pre-training method of the previous section, but larger for the negative spin case (virtual 2) and smaller for virtual 1 and positive spins.

\begin{table}[t] \centering 
\caption{Monte Carlo derived confidence levels (1$\sigma$).}
\label{tab:mocks_tier_grid_3x3_absolute_2sigma} 
\begin{tabular}{llcrrr} 
\toprule 
Param. & Exp. & Virtual & EHT & ngEHT & BHEX \\
\midrule
$\alpha_{\ast}$ & I &  1 & 0.31 & 0.21 & 0.05 \\
& I &  2 & 0.44 & 0.31 & 0.08 \\
& II &  1 & 0.13 & 0.10 & 0.02 \\
& II &  2 & 0.28 & 0.20 & 0.05 \\
& IV &  1 & 0.25 & 0.19 & 0.04 \\
& IV &  2 & 0.27 & 0.24 & 0.3 \\
\midrule
$i[^{\circ}]$ & II &  1 & 9.3 & 8.9 & 2.3 \\
& II &  2 & 7.6 & 4.9 & 1.0 \\
& IV &  1 & 7.9 & 8.2 & 3.8 \\
& IV &  2 & 5.8 & 4.8 & 1.9 \\
\midrule
$\theta_{z}[^{\circ}]$& IV &  1 & 14 & 14 & 8.6 \\
& IV &  2 & 13 & 12 & 11 \\
\bottomrule
\end{tabular} 
\tablefoot{Confidence levels from the Monte Carlo inference runs for two virtual observations and three arrays.}
\end{table}

\section{ Discussion }
\label{sec:discussion}

There are several insights worth discussing further in the presented results.
In the case of equatorial hot spots, the extreme accuracy of \ramiland predictions solidifies the existence of a non-degenerate underlying relation between $\Delta PA$ and hot-spot parameters, including $\alpha_*$ and $i$; suggesting hot-spot phenomenology can be an excellent mechanism for extracting properties of the near vicinity of super-massive black holes. 
Furthermore, our research reinforces the validity of time-averaged $\dpa$ as a proxy for the $i=0$ case across a variety of actual inclinations ($i\leq60^{\circ}$), consistent with prior research \citep{Walia25, Y26}. This shows how pure the connection of $\Delta PA$ to $\alpha_*$ is in the simplest case.

\begin{figure*}[h]
   \includegraphics[width=0.902\textwidth]{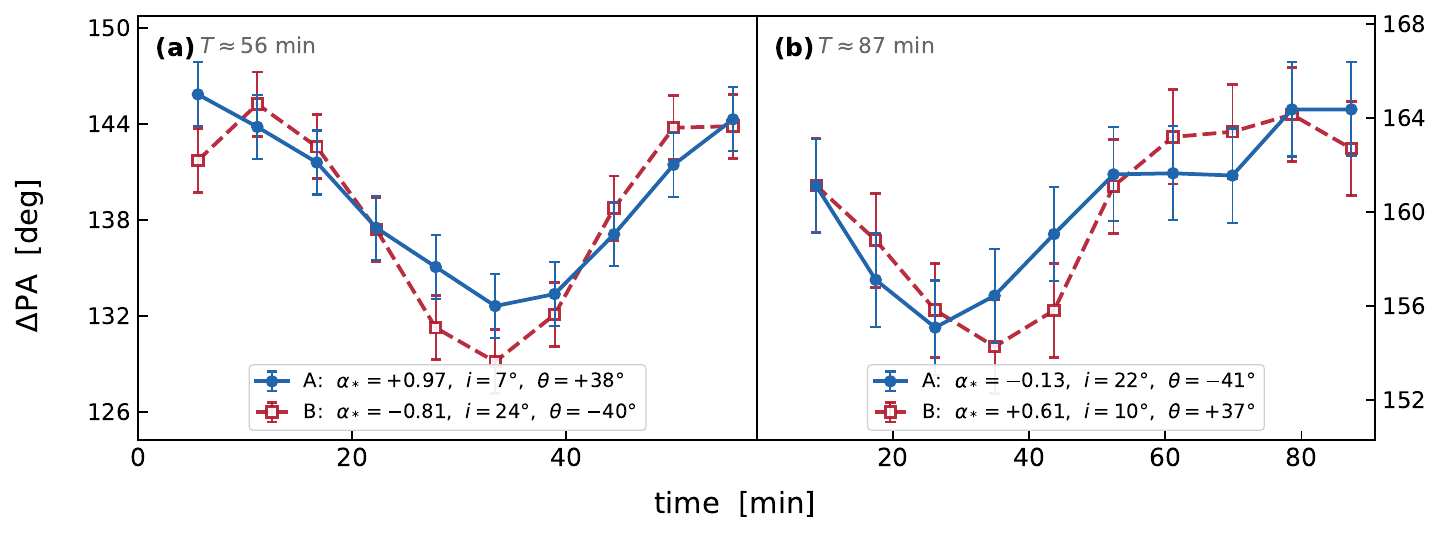}
   \caption{Different parameter configurations leading to similar $\Delta PA$ curves within BHEX estimated error bars.}
   \label{fig:exp4_degens}
\end{figure*}

In the case of non-equatorial hot spots, we encounter another interesting effect that remains understudied. That is the very close degeneracy between $\theta_z$, $i$ and $\alpha_*$. From the whole Section \ref{sec:results} and especially \ref{sec:Mock}, it is clear that there are right combinations that can reproduce near-identical $\dpa$ curves. To demonstrate this even further, we present Figure \ref{fig:exp4_degens} where we show two panels of near-identical $\Delta PA$ curves, or at least indistinguishable under BHEX error levels, produced from different models. In both cases, $\alpha_*$ and $\theta_z$ signs are reversed, with $i$ values switching from medium to low, to accommodate the similar shape.

\begin{figure*}
    \centering
    \includegraphics[width=0.902\linewidth]{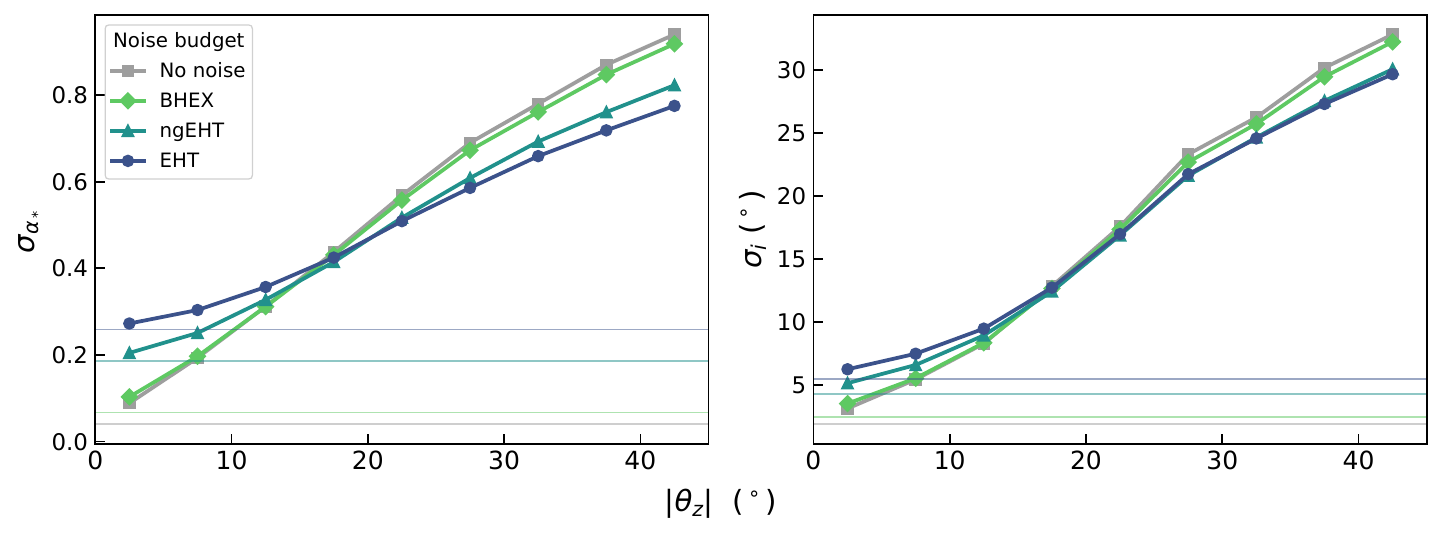}
    \caption{Resulting $\sigma_{\alpha_*},\, \sigma_i$, evaluated on the test set of Experiment IV: non-equatorial orbits, when inferred with Experiment II, trained on equatorial orbits. The horizontal lines denote $\sigma$ values estimated from Experiment II.}
    \label{fig:exp6}
\end{figure*}

This may appear to be a drawback of our method at first, but the finding itself is more valuable. This is because if we could acquire prior information on either $i$ or $\theta_z$, the degeneracy would break, and we should expect spin estimates to approach the Exp II levels. For $i$ that could be achieved by applying the numerous estimations for SgrA*'s inclination \citep[$i\sim150-170^{\circ}$, e.g.][]{gravityMichi,eht:2022_paperV,yfantis24a}. In the case of $\theta_z$ there are only softer indications for non-equatorial orbits presently \citep[e.g. from super-Keplerian velocities,][]{aimar_2023_plasmoid,yfantis24b,antonop025_spiral}, but in the future, and given movie observations, there could arise new ways, for example from testing the relative deformation of the hot spot throughout a period, that seems to be intensified on high $\theta_z$ values, and nullified in negative values.

To demonstrate the importance of Experiments IV and V even further we perform another test; namely the evaluation of prediction errors of Exp. II, while using $\Delta PA$ curves from non-equatorial orbits of increasingly higher $\theta_z$. The results can be seen in Figure \ref{fig:exp6} for $\alpha_*$ and $i$ respectively. For both parameters prediction errors rise rapidly with the increase of $\theta_z$, especially in the case of $i$. Interestingly, already at $\theta_z \sim 10^\circ$ both $\sigma_{\alpha_*},\,\sigma_i$ have more than tripled in the no-noise and BHEX cases, highlighting the sensitivity on $\theta_z$.

Lastly, another surprising result was the plateau in the errors of partial orbits (Exp. III, V) down to $50\%$ of the orbit. This provides an avenue for meaningful estimates from the first observations, which will most likely be less detailed, noisier, and possibly only for a fraction of a period, given the limited dynamic range compared to later missions.

Naturally, there may be effects of hot-spot phenomenology that we have not accounted for. Most notably, spiraling spots \citep[e.g.][]{Ruales2026_spiraling}, or other types of non-circular or tilted orbits. Additionally, many effects can obscure observational feasibility. Recall that our method applies only once the $\Delta PA$ has been extracted. We have indicated some generic expected errors in \ref{sec:Mock}, but other effects can occur. Most obviously, the existence of two bright spots on the screen mimicking a secondary image. A clear differentiating feature can be the time difference in appearance between the two features, precisely calculated in our simulations. 

Another possibility is the relative total intensity between the two features expected around $5-10\%$ in the general case. This comes with a problem: the available dynamic range of the imaging algorithm; in essence, the ratio of the brightest to the dimmest pixel in the image. Current algorithms are expected to have a dynamic range around $8-10$, so they are at the limit for covering the $5-10\%$ ratio of lensed emission. But since the lensed photons continue to travel through the disk and overlap with the disk's primary emission, the pixels would not appear at $5\%$ of the hot spot intensity, but well above it, depending on the disk's intensity. Furthermore, the primary emission in vertical magnetic fields dims significantly at a certain phase due to relativistic aberration \citep{Narayan21_aberation,Vos:2022}, making the secondary image brighter for roughly a third of the orbit \citep[appendix in][]{yfantis24a}, creating a clear avenue for identifying a disappearing hot spot with a secondary image.  

These are all excellent concerns, but beyond the scope of this project, which focuses on possible parameter estimation after an observation has extracted the required parameters $\dpa,\,\rs$, and $P$. The observational extraction is a natural and compelling next step for future studies. While a great alternative avenue for extraction would be to train STIHOS on synthetic observations of the VLBI data during the hot-spot orbit, thereby excluding any errors accumulated during imaging.
Furthermore, here we have discussed the case of a single observation, but since $\alpha_*,\,i$ are not expected to change on meaningful timescales, the accuracy could increase drastically simply by stacking observations from different flaring events.

Regardless of the method's current accuracy and feasibility, the value of this study extends beyond spin estimation. The findings complement theoretical works on photon rings \citep[e.g.][]{Gralla2019,Johnson2020} and hot spots \citep{Gralla2020_lensing,Prashant2024} in a simple manner, bypassing cumbersome mathematical formalism and focusing on the more realistic n=1 emission.

\section{Summary}
\label{sec:conclusion}

We have created and trained \ramiland, a DL algorithm for estimating the spin and inclination of black holes using secondary images from hot-spot observations, which is fully available at \link. The algorithm works exceptionally well for equatorial spots under the non-error assumption ($\sigma_{\alpha_*}=0.04,\,\sigma_i=2^{\circ}$). Including non-equatorial hot spots without prior information can, in some cases, create unbreakable degeneracies. For equatorial spots, current ground arrays yield constraints of $\sigma_{\alpha_*} \sim 0.3$ and $\sigma_{i} = 10^{\circ}$, while positive spins seem to be better constrained. In the case of space interferometry, like BHEX, $2\sigma_{\alpha_*}$ drops below 0.1, constraining spin with unprecedented precision. 

Looking forward, the proof of the photon ring's existence appears to be an increasingly realistic goal for the community of relativity, strong gravity, and very long baseline interferometry. In comparison, observing a hot spot together with its secondary image requires less angular resolution, provided the observation timing is favorable. Thus, we are confident that this research will be fruitful in the near future and hopeful that further work will refine the methods discussed here.

\section*{Data availability}
Simulation data can be shared upon reasonable request. 

\begin{acknowledgements}

	We thank M. Foschi for useful comments. We also thank M. Wielgus and J. L. Gomez's group at IAA for useful discussions. AIY acknowledges funding from the European Union’s Horizon Europe research and innovation program under grant agreement No. 101093934 (RADIOBLOCKS). Also, support from the COMA cluster at Radboud University, where a large part of the simulations was performed. Special thanks to DTU Compute and the supercomputing cluster Titans, where the compute cluster where the DL experiments were run (\url{https://titans.compute.dtu.dk/}). Software used in the paper: Python, \ipole, \ramiland (available in \link).
	
\end{acknowledgements}
 
\bibliographystyle{aa} 
\bibliography{library}

\end{document}